\documentclass[showpacs,aps,prl,twocolumn,superscriptaddress]{revtex4}
\usepackage{graphicx} 
\usepackage{dcolumn}
\usepackage{bm}

\begin{document}

\title{Magneto-spectroscopy of Highly-Aligned Carbon Nanotubes:\\Identifying the Role of Threading Magnetic Flux}

\author{J.~Shaver}
\affiliation{Electrical and Computer Engineering Department, Rice
University, Houston, Texas 77005, USA}

\author{S.~A.~Crooker}
\affiliation{National High Magnetic Field Laboratory, Los Alamos,
New Mexico 87545, USA}

\author{J.~A.~Fagan}
\affiliation{National Institute of Standards and Technology,
Gaithersburg, Maryland 20899, USA}

\author{E.~K.~Hobbie}
\affiliation{National Institute of Standards and Technology,
Gaithersburg, Maryland 20899, USA}



\author{N.~Ubrig}
\affiliation{Laboratoire National des Champs Magn\'{e}tiques
Puls\'{e}s, 31400 Toulouse, France}

\author{O.~Portugall}
\affiliation{Laboratoire National des Champs Magn\'{e}tiques
Puls\'{e}s, 31400 Toulouse, France}

\author{V.~Perebeinos}
\affiliation{IBM Research Division, T. J. Watson Research Center, Yorktown Heights, New York 10598, USA}

\author{Ph.~Avouris}
\affiliation{IBM Research Division, T. J. Watson Research Center,
Yorktown Heights, New York 10598, USA}

\author{J.~Kono}
\email[]{kono@rice.edu}
\affiliation{Electrical and Computer Engineering Department, Rice University, Houston, Texas 77005, USA}


\date{\today}

\begin{abstract}
We have investigated excitons in highly-aligned single-walled carbon nanotubes (SWCNTs) through optical spectroscopy at low temperature ($1.5$~K) and high magnetic fields ($\textbf{\textit{B}}$) up to 55~T.  SWCNT/polyacrylic acid films were stretched, giving SWCNTs that are highly aligned along the direction of stretch ($\hat{n}$).  Utilizing two well-defined measurement geometries, $\hat{n}\parallel\textbf{\textit{B}}$ and $\hat{n}\perp\textbf{\textit{B}}$, we provide unambiguous evidence that the photoluminescence energy and intensity are only sensitive to the $\textbf{\textit{B}}$-component parallel to the tube axis.  A theoretical model of one-dimensional magneto-excitons, based on exchange-split `bright' and `dark' exciton bands with Aharonov-Bohm-phase-dependent energies, masses, and oscillator strengths, successfully reproduces our observations and allows determination of the splitting between the two bands as $\sim4.8$~meV for (6,5) SWCNTs.
\end{abstract}


\maketitle


Elementary excitations in matter are strongly influenced by Coulomb interactions, especially when they are quantum-confined in low-dimensional structures~\cite{CB,1D}. Correlated electron-hole pairs (or excitons) in one-dimensional (1-D) systems are known to behave very differently from those in higher dimensions~\cite{Loudon59AJP,ElliotLoudon5960JPCS,OgawaTakagahara91both,Citrin92PRL}.  Recently, individualized single-walled carbon nanotubes (SWCNTs)~\cite{OconnelletAl02Science} have emerged as an ideal system for investigating excitons in the ultimate limit of 1-D quantum confinement and relatively small amounts of disorder.  The existence of two equivalent energy valleys in momentum space combined with strong Coulomb interactions lead to a splitting between `dark' and `bright' excitonic states in the lowest-energy spin-singlet
manifold~\cite{PerebeinosetAl04PRL,ZhaoMazumdar04PRL,SpataruetAl05PRL,PerebeinosetAl05NL,ChangetAl06cond-mat,Ando06JPSJ,ShaveretAl07NL,ShaverKono07LPR}.
Magnetic fields ($\textit{\textbf{B}}$) offer a controllable way to mix dark and bright wavefunctions and redistribute oscillator strengths~\cite{Ando06JPSJ,ShaveretAl07NL,ShaverKono07LPR}.  In existing magneto-optical studies on
liquid~\cite{ZaricetAl04Science,ZaricetAl04NL,ZaricetAl06PRL} and solid~\cite{ShaveretAl07NL,MortimerNicholas07PRL,MortimerNicholas07PRB} samples it is impossible to isolate the effects of each field component due, respectively, to the dynamic and isotropic natures of the samples used.

\begin{figure}[t]
 \includegraphics[scale=.85]{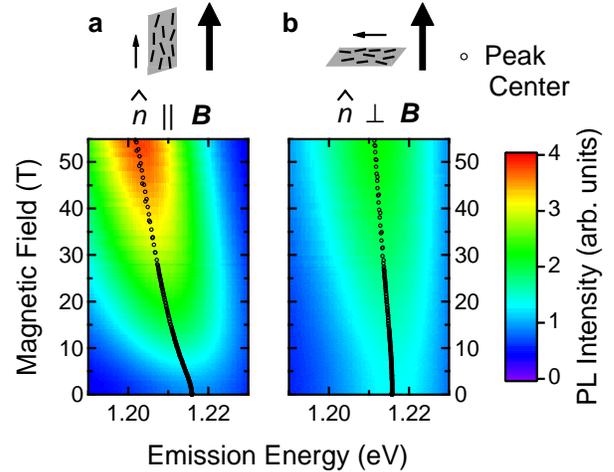}\\
 \caption{(color online). Contour plots of PL intensity as a function of magnetic field with nanotubes aligned (a) parallel ($\hat{n}\parallel\textbf{\textit{B}}$) to and (b) perpendicular ($\hat{n}\perp\textbf{\textit{B}}$) to the magnetic field.  As the magnetic field increases, the PL peak shifts ($\circ$ markers) and increases in intensity.  The magnitude of the shift and increase is much greater for $\hat{n}\parallel\textbf{\textit{B}}$ than $\hat{n}\perp\textbf{\textit{B}}$.} \label{PL_contour}
\end{figure}

Utilizing \emph{highly-aligned} and \emph{individualized} SWCNTs in magneto-optical experiments, we provide unambiguous evidence that the photoluminescence (PL) emission energy and intensity are \emph{only} affected by a tube-threading, \emph{parallel}, magnetic field ($B_\parallel$).  We performed low-temperature (1.5~K) PL and absorption spectroscopy of stretch-aligned SWCNT films in $\textbf{\textit{B}}$ up to 55~T in two well-defined geometries.  In the Voigt geometry,
$\hat{n}\parallel\textbf{\textit{B}}\perp\textbf{\textit{k}}$ (where $\hat{n}$ is a unit vector in the alignment direction and $\textbf{\textit{k}}$ is the light propagation vector), we observed large PL energy shifts and intensity increases with $\textbf{\textit{B}}$, while in the Faraday geometry, $\hat{n}\perp\textbf{\textit{B}}\parallel\textbf{\textit{k}}$, we observed small PL energy shifts and intensity increases.  Notably, we find excellent agreement when scaling our data to the SWCNT-parallel component of $\textbf{\textit{B}}$ ($B_\parallel$), conclusively demonstrating that a tube-threading magnetic field is the responsible component.  We explain these changes with our model of 1-D magneto-excitonic bands~\cite{ShaveretAl07NL,ShaverKono07LPR} based on the Aharonov-Bohm
effect~\cite{AjikiAndo93JPSJ,AjikiAndo94Physica,ZaricetAl04Science,KonoRoche06CRC}.


PL measurements were performed at the National High Magnetic Field Laboratory (NHMFL) in Los Alamos, using the 60~T long (2.5~s, programmable) pulse magnet powered by a 1.4~GVA motor-generator.  We used resonant $E_{22}$ excitation of (6,5) tubes with a dye laser tuned to 570~nm (or 2.175~eV).  Absorption was carried out at the Laboratoire National des Champs Magn\'{e}tiques Puls\'{e}s (LNCMP) in Toulouse with a capacitor-bank-driven, 60~T short (150~ms) pulse magnet.  A quartz-tungsten-halogen lamp was used for absorption measurements.  Signals were dispersed on 300-mm monochromators and recorded with InGaAs diode arrays using a typical exposure-plus-readout time of 1-2~ms.

The samples were DNA-wrapped CoMoCAT SWCNTs dispersed in a polyacrylic acid (PAA) matrix.  These samples have strong PL signal, sharp absorption features, are temperature stable, and can be stretch-aligned~\cite{FaganetAl07PRL}.  As the PAA film is stretched, the embedded nanotubes align to the direction of pull ($\hat{n}$).  The degree of alignment is characterized by the dimensionless nematic order parameter $S=(3\left<\cos^2{\theta}\right>-1)/2$, which scales from 0 (for random orientation) to 1 (for uniaxial orientation), where $\theta$ the angle between $\hat{n}$ and an individual
SWCNT~\cite{Hobbie04JCP}.  For large values of $S$ a single effective angular value ($\theta_{\rm eff}$) characterizes the sample and is defined through $\cos^2{\theta_{\rm eff}} \equiv \left<\cos^2{\theta}\right> = (2S+1)/3$~\cite{dichroism}.  The films used in this study typically had $S\sim 0.8$, determined by polarized Raman spectroscopy, which gives $\theta_{\rm eff}\approx 21^\circ$~\cite{FaganetAl07PRL}.

Figure~\ref{PL_contour} shows a surface plot of PL vs.~$B$ for (a) $\hat{n} \parallel \textbf{\textit{B}}$ and (b) $\hat{n} \perp \textbf{\textit{B}}$.  At high $B$, the dark exciton state is brightened, increasing the PL intensity and red-shifting the emission energy.  For $\hat{n} \parallel \textbf{\textit{B}}$, the amounts of brightening and red-shift are dramatically larger than for $\hat{n} \perp \textbf{\textit{B}}$.  These data are in contrast to
recent reports of a substantial magnetic brightening of randomly-oriented frozen samples in the Faraday geometry ($\textbf{\textit{k}} \parallel \textbf{\textit{B}}$)~\cite{MortimerNicholas07PRL,MortimerNicholas07PRB}.  Brightening in randomly-oriented tubes, however, is likely due to the excitation of (and detection of PL from) the significant SWCNT population having an appreciable projection along $\textbf{\textit{B}}$ (e.g., consider a tube at $45^\circ$ to $\textbf{\textit{B}}$).  The use of highly-aligned SWCNTs in this study markedly reduces the population of SWCNTs oriented with any projection along $\textbf{\textit{B}}$, allowing us to separate the influence of the parallel ($B_\parallel$) from the perpendicular
field component ($B_\perp$).

\begin{figure}
 \includegraphics[scale=0.88]{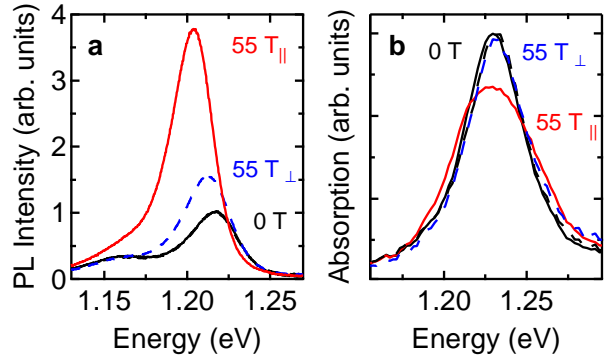}
 \caption{(color online). (a) PL spectra at 0~T and 55~T with $\hat{n} \parallel \textbf{\textit{B}}$ and $\hat{n} \perp \textbf{\textit{B}}$.  At zero field, both configurations have the same spectra (black solid).  At high field, the intensity is increased and emission energy red-shifted for $\hat{n} \parallel \textbf{\textit{B}}$ (red solid) much more than for $\hat{n} \perp \textbf{\textit{B}}$ (blue dashed).  (b) Absorption spectra at 0~T and 55~T in the same configurations.  The intensity is decreased and the peak broadened (integrated oscillator strength is conserved) for $\hat{n} \parallel \textbf{\textit{B}}$ (red solid) while for $\hat{n} \perp \textbf{\textit{B}}$ (blue dashed) they are unchanged at high field.} \label{Spectra}
\end{figure}

Figure~\ref{Spectra}(a) shows ``slices'' of Fig.~\ref{PL_contour} at 0~T and 55~T. The PL intensities are normalized to zero field where the spectra for both configurations are identical.  It is clearly seen that $\hat{n} \parallel \textbf{\textit{B}}$ demonstrates stronger brightening and larger red-shifts at a given $B$ than $\hat{n} \perp \textbf{\textit{B}}$. Figure \ref{Spectra}(b) shows absorption spectra at 0~T and 55~T in the same configurations.  As with PL, changes in absorption spectra are much more distinct when $\hat{n} \parallel \textbf{\textit{B}}$.

%

\begin{figure}
 \includegraphics[scale=.88]{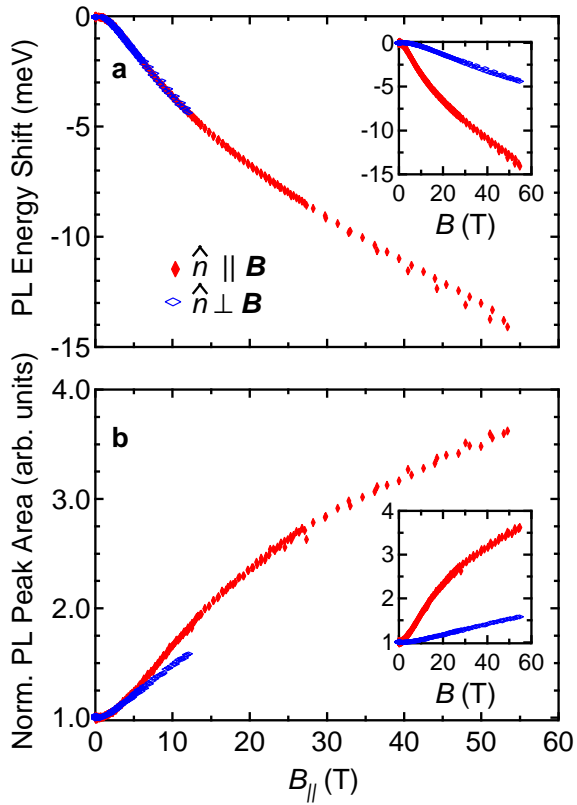}\\
 \caption{(colo online). PL peak shift (a) and normalized peak area (b) as a function of the SWCNT-parallel component of $\textbf{\textit{B}}$ ($\textit{B}_\parallel$) for $\hat{n} \parallel \textbf{\textit{B}}$ and $\hat{n} \perp \textbf{\textit{B}}$. The peak energy matches exactly for both geometries.  The peak area matches very well, except for a slight mismatch at high fields.  Insets are the same values as a function of total field strength ($B$).} \label{PL_analysis}
\end{figure}

We extracted the PL peak energy and integrated area by fitting spectral slices from Fig.~\ref{PL_contour} at each $B$ with Lorentzian lineshapes using a modified Levenberg-Marquardt algorithm. The insets of Fig.~\ref{PL_analysis}(a) and \ref{PL_analysis}(b) show the $B$-dependent PL peak energy shift and normalized area vs.~$B$.  At high field the PL intensity of $\hat{n} \parallel \textbf{\textit{B}}$ is $>3.5$ times larger and red-shifted by $\sim 15$~meV while the intensity of $\hat{n} \perp \textbf{\textit{B}}$ is only $\sim1.5$ times larger and red-shifted by $\sim 5$~meV.  The main figures~\ref{PL_analysis}(a) and \ref{PL_analysis}(b) will be
discussed later.  It is also clear from Fig.~\ref{PL_analysis}(a) that the $B$-dependence of the peak energy is \emph{not} linear as expected from $B$-dependent band gap theory~\cite{AjikiAndo93JPSJ}.  This nonlinear $B$-dependence is a direct result of the excitonic nature of and balance of oscillator strengths between the dark and bright states~\cite{Ando97JPSJ,Ando06JPSJ} and can be quantitatively explained as follows.

\begin{figure}
 \includegraphics[scale=.8]{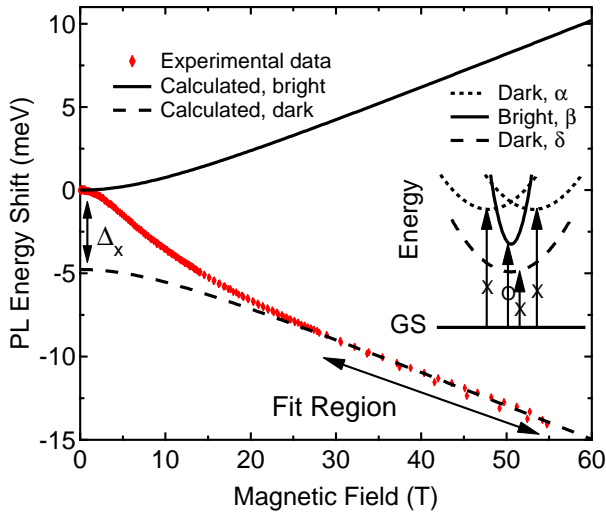}\\
 \caption{(color online). Experimental (red diamonds) and calculated (solid and dashed lines) peak energy as a function SWCNT-parallel component of magnetic field. Values of $\Delta_x\sim4.8$~meV and $\mu=0.93~\textrm{meV/T-nm}^2$ were determined from fitting the high-field region of the experimental data and used for calculation of the $K = 0$ dark magneto-exciton energy, $\epsilon_\delta(B)$ (dashed), and the $K = 0$ bright magneto-exciton energy, $\epsilon_\beta(B)$ (solid). The inset is an energy level diagram, showing the dispersions of the four lowest-energy singlet excitons with allowed (circle) and prohibited (crosses) transitions indicated.  GS: ground state.\label{Peak_pos_fit}}
\end{figure}

The doubly-degenerate valence and conduction bands in SWCNTs give rise to four spin-singlet excitons (see Fig.~\ref{Peak_pos_fit} for an energy level diagram).  To be optically active with linear polarization, the angular momentum must be zero and the excitonic wavefunction must have odd spatial symmetry. The lowest-energy exciton has zero angular momentum and even spatial symmetry, and thus, it cannot decay radiatively (i.e., it is ``dark''). The odd
symmetry, zero angular momentum exciton state has a slightly higher energy due to the Coulomb exchange interaction and can decay radiatively (i.e., it is ``bright'').  The remaining two exciton states, with finite angular momentum, cannot decay radiatively and are predicted to be higher in energy. The hyperbolic energy dispersions~\cite{MintmireWhite98PRL} of the zero angular momentum dark ($i = \delta$) and bright ($i = \beta$) excitons can be written as by $E_i (K) = \sqrt{E_i(0)^2 + E_i(0) \hbar^2 K^2/m_i}$, where $E_i(0)$ is the energy at the bottom of the band [$E_\beta (0)
E_\delta (0)$], $m_i$ is the exciton effective mass, and $K$ is the wavevector associated with the exciton center-of-mass momentum.

A magnetic field breaks time reversal symmetry and mixes the zero angular momentum dark and bright exciton wavefunctions.  This mixing redistributes the oscillator strength between the two excitons and modifies their effective masses (discussed elsewhere~\cite{ShaverKono07LPR,ShaveretAl07NL}), which leads to the brightening of the lowest-energy dark exciton.  At non-zero $B$, these states are mixed by an Aharonov-Bohm term, $\Delta_{\rm AB} = \mu\phi$, where $\mu$ is a proportionality constant and $\phi = \pi B_\parallel d^2/4$ is the flux threading a SWCNT with diameter $d$.

We use the following Hamiltonian to describe this two-band 1-D magneto-exciton system:
\begin{equation}
\footnotesize \hat{H}(K) = \frac{E_\beta(K)+E_\delta(K)}{2} \hat{I}
+ \frac{E_\beta(K)-E_\delta(K)}{2} \hat{\sigma_z} +
\frac{\Delta_{\rm AB}(B)}{2} \hat{\sigma_x} \label{hamiltonian}
\end{equation}
\noindent where $\hat{I}$ is a unit matrix and $\hat{\sigma_x}$ and $\hat{\sigma_z}$ are Pauli matrices.  The eigenvalues of the Hamiltonian [Eq.~(\ref{hamiltonian})] are
\begin{equation}
\footnotesize \varepsilon_{\delta,\beta}(K,B) = \frac{E_\beta(K)+E_\delta(K) \mp \sqrt{\left\{E_\beta(K)-E_\delta(K)\right\}^2 + \Delta_{\rm AB}^2(B)}}{2}, \label{dispersion1}
\end{equation}
\noindent where $\varepsilon_\beta (K,B) > \varepsilon_\delta (K,B)$.  The higher energy, finite angular momentum exciton bands ($i=\alpha$) are predicted to have similar masses to the dark band and have negligible magnetic field
dependence~\cite{PerebeinosetAl05NL,Ando06JPSJ}.


The $\textit{B}$-dependent term in the Hamiltonian [Eq.~(\ref{hamiltonian})] depends on the flux $\phi$ that threads the nanotube, which will depend on the angle $\theta$ it makes with $\textit{\textbf{B}}$. Thus, the relevant component of $\textit{\textbf{B}}$ for the macroscopic sample is $B_\parallel = \textit{B}\cos{\theta_{\rm eff}}$ for $\hat{n}\parallel\textbf{\textit{B}}$ and $B_\parallel = \textit{B}\sin{\theta_{\rm eff}}$ for $\hat{n}\perp\textbf{\textit{B}}$.  To demonstrate that only $B_\parallel$ is important in determining the PL peak energy, we plotted the peak energy data for both configurations in Fig.~\ref{PL_analysis}(a) as a function of $B_\parallel$. Using $\theta_{\rm eff}\approx13^\circ$ results in a nearly perfect overlap in functional form of peak shift vs.~$\textit{B}_\parallel$ between the two configurations and gives $S\approx0.9$.  This is a slightly higher alignment than determined from polarized Raman data~\cite{FaganetAl07PRL}, possibly due to the contribution of non-luminescent species such as highly bent tubes to the Raman experiment. The PL peak area vs.~$\textit{B}_\parallel$ in Fig.~\ref{PL_analysis}(b) is scaled using the same $\theta_{\rm eff}$.  It also agrees quite well, except a slight mismatch at high magnetic fields.  This scaling shows that observed magnetic field dependence is due to $B_\parallel$ and that $B_\perp$ has minimal impact on the PL emission properties of 1D excitons in SWCNTs. Large PL enhancements recently reported for randomly aligned samples in the Faraday geometry are therefore most likely due to a population of SWCNTs with significant threading flux. This emphasizes the need for highly aligned samples in accurately determining the effects of magnetic field components and measurement geometry in SWCNT magneto-PL experiments.

In previous magneto-PL experiments at room temperature, the PL energy was centered on the bright state at zero $B$, and as $B$ increases, so does the oscillator strength of the dark state, broadening and eventually splitting the PL peak at very high $B$~\cite{ZaricetAl06PRL}.  In the current experiment, which was performed at low temperature ($T$), the PL peak energy is also centered on the bright state at zero $B$.  However, in contrast to the room temperature experiment, as $B$ increases such that $\Delta_{\rm AB}\gg k_{\rm B} T$, where $k_{\rm B}$ is the Boltzmann constant, the exciton population is restricted to the formerly dark, low-energy state, and therefore, the $B$-dependence of the peak energy tracks $\varepsilon_\delta (K,B)$.  As only excitons with momentum near $K=0$ can decay radiatively, we simplify Eq.~(\ref{dispersion1}) to $\epsilon_{\delta,\beta}(B) = \left(-\Delta_x \mp \sqrt{\Delta_x^2 + \Delta_{\rm AB}^2(B)}\right)/2$, where $\Delta_x = E_{\beta}(0) - E_{\delta}(0)$ is the dark-bright energy splitting in zero $B$. Note that as the zero-field PL emission is only from the bright state, we set $E_\beta(0)=0$ to plot the energy shift from zero field vs.~$B$. We can fit our high field energy shift data (indicated in Fig.~\ref{Peak_pos_fit}), extracting $\Delta_x\sim4.8$~meV and $\mu=0.93~\textrm{meV/T-nm}^2$. These values are then used to extrapolate the total field dependence
$\epsilon_\delta (B)$ and $\epsilon_\beta (B)$. Figure~\ref{Peak_pos_fit} shows experimental PL energy from $\hat{n}\parallel\textbf{\textit{B}}$ and calculated energies from $\epsilon_{\delta,\beta}(B)$. The excellent agreement of calculated and experimental peak energies strongly supports our model and conclusively shows the significance of $B_\parallel$ in determining magneto-optical properties of SWCNTs.

For the first time in SWCNT magneto-optical experiments, we utilized highly aligned samples to demonstrate, conclusively, that a parallel, tube-threading magnetic field is the dominant source of changes in spectral energy and intensity at high values of $\textbf{\textit{B}}$, and determined the splitting between dark and bright bands of (6,5) SWCNTs in our  sample.  This is supported by the striking overlap of PL energy shift as a function of $B_\parallel$ in different measurement geometries and the accompanying fit to our model based on exchange-split bright and dark exciton bands with Aharonov-Bohm-phase-dependent energies, masses, and oscillator strengths.

\begin{acknowledgments}
This work was supported by the Robert A.~Welch Foundation (through grant No.~C-1509) and the National Science Foundation (through Grants Nos. DMR-0134058, DMR-0325474, and OISE-0437342). We thank Ajit Srivastava for helpful discussions and the support staff of the Rice Machine Shop, NHMFL, and LNCMP.
\end{acknowledgments}


\end{document}